\documentclass[twocolumn,prl,aps]{revtex4-1}
\usepackage[T1]{fontenc}
\usepackage[latin9]{inputenc}
\usepackage{bm}
\usepackage{amsmath}
\usepackage{amssymb}
\usepackage{graphicx}
\usepackage{esint}
\usepackage[unicode=true,
 bookmarks=false,
 breaklinks=false,pdfborder={0 0 1},backref=false,colorlinks=false]
 {hyperref}

\makeatletter

\usepackage{color}
\usepackage{bm}
\usepackage{bbold}
\usepackage{soul}
\usepackage{epsfig}
\usepackage{verbatim}
\usepackage{array}

\newcommand{\be}{\begin{equation}}
\newcommand{\ee}{\end{equation}}
\newcommand{\bea}{\begin{eqnarray}}
\newcommand{\eea}{\end{eqnarray}}

\makeatother

\begin{document}
\title{Non-Abelian Geometric Dephasing}
\author{Kyrylo Snizhko,$^{1}$ Reinhold Egger,$^{2}$ and Yuval Gefen$^{1}$}
\affiliation{$^{1}$~Department of Condensed Matter Physics, Weizmann Institute,
Rehovot, Israel ~\\
 $^{2}$~Institut f\"ur Theoretische Physik, Heinrich-Heine-Universit\"at,
D-40225 D\"usseldorf, Germany}
\date{\today}
\begin{abstract}
We study the adiabatic dynamics of degenerate quantum states induced
by loop paths in a control parameter space. The latter correspond
to noisy trajectories if the system is weakly coupled to environmental
modes. On top of conventional dynamic dephasing, we find a universal
non-Abelian geometric dephasing (NAGD) contribution and express it
in terms of the non-Abelian Berry connection and curvature. We show
that NAGD implies either decay or amplification of coherences as compared to the coherences when only dynamic dephasing is present.
The full NAGD matrix structure can be probed through interference
experiments. We outline such a detection scheme for modified Majorana
braiding setups.
\end{abstract}
\maketitle
\emph{Introduction.---}Ever since the formulation of the Standard
Model, non-Abelian gauge theories have played a central role in modern
physics. In particular, concepts like the non-Abelian Berry connection
and curvature \citep{Nakahara} have been of key importance to such
diverse topics as wave function dynamics in degenerate quantum systems
\citep{Wilczek1984}, the fractional quantum Hall effect \citep{Arovas1984,Wen1991},
topological \citep{Nayak2008} and geometric \citep{Pachos1999,Jones2000}
quantum computation, nuclear quadrupole resonance \citep{Zee1988},
topological insulators \citep{Hasan2010,Yang2014} and topologically
ordered phases \citep{Wen2017}, degenerate Bloch bands in solids
\citep{Xiao2010}, synthetic non-Abelian gauge fields in ultracold
atom systems \citep{Osterloh2005,Li2016}, or for a geometric understanding
of Christoffel symbols in general relativity \citep{Carroll}. In
particular, the non-Abelian theory generalizes the notion of adiabatic
quantum transport from the non-degenerate Abelian case to $N$-fold
degenerate state spaces with $N\ge2$. In the Abelian case, a state
picks up the celebrated geometric Berry phase when the time dependence
of the Hamiltonian stems from adiabatically traversing a closed loop
in parameter space \citep{Berry1984}. For $N\ge2$, the Berry phase
factor is replaced by a unitary $N\times N$ Berry matrix ${\cal U}_{B}$,
which again only depends on the geometry of the parameter loop \citep{Wilczek1984}.

It is of both fundamental and applied interest to understand what
happens when the control parameters are subject to random fluctuations
(noise) due to the system being weakly coupled to environmental modes.
In the Abelian case, this problem has been extensively studied \citep{Ellinas1989,Gamliel1989,Gaitan1998,Avron1998,Carollo2003,DeChiara2003,Whitney2003,Whitney2005,Syzranov2008}.
Most importantly, for a dissipative spin-$1/2$ system subject to
a cyclic magnetic field trajectory, Refs.~\citep{Whitney2003,Whitney2005}
have predicted geometric dephasing contributions that depend on the
sign of the winding (the trajectory orientation) along a closed path.
This prediction has recently been confirmed \citep{Berger2015} in
superconducting nanocircuit experiments \citep{Leek2007}. We here
establish a theoretical framework for studying the non-Abelian adiabatic
dynamics of \emph{open} quantum systems. In particular, we show that
universal geometric dephasing, described by a model-independent expression,
is also present in the non-Abelian case. The nontrivial matrix structure
associated with NAGD is responsible for richer physics and causes
characteristic and experimentally observable differences compared
to the Abelian counterpart. Experimental protocols for observing NAGD
and confirming its distinctive features are given below, see also
Ref.~\citep{prblong} for additional details. We illustrate our ideas
for noisy Majorana braiding setups, where braiding is executed by
running time-dependent protocols for the tunnel matrix elements between
different pairs of Majorana bound states. Majorana braiding protocols
are currently also of considerable experimental interest \citep{Alicea2012,Lutchyn2017}.
From a general perspective, a thorough understanding of dephasing
mechanisms in systems with degenerate subspaces has to include the
NAGD contributions discussed here.

\emph{Physical origin of NAGD.---}Before presenting explicit expressions
for NAGD, we first qualitatively explain its essence. For a non-degenerate
system, on top of the dynamic phase, $e^{i\varphi_{d}}=e^{-i\int dtE(t)}$,
the adiabatic evolution of a state in the Hilbert space leads to a
Berry phase factor $e^{i\varphi_{B}}$ \citep{Berry1984}. For an
open system, both the state trajectory and the energy fluctuate due
to the influence of the environment. Upon averaging over such fluctuations,
the phase $e^{i\varphi_{d}+i\varphi_{B}}$ will be replaced by $e^{i\varphi_{d}-\Gamma_{\mathrm{dyn}}T}e^{i\tilde{\varphi}_{B}-\Gamma_{g}}$
\citep{Whitney2003,Whitney2005}. With the time duration $T$ of the
protocol, the dynamic dephasing term, $\Gamma_{\mathrm{dyn}}T$, comes
from fluctuations of the dynamic phase. Geometric dephasing is encoded
by $\Gamma_{g}$, and there is a correction $\varphi_{B}\to\tilde{\varphi}_{B}$.
Executing the same protocol in the opposite direction, we have $\tilde{\varphi}_{B}\rightarrow-\tilde{\varphi}_{B}$
but also $\Gamma_{g}\rightarrow-\Gamma_{g}$ \citep{Whitney2003,Whitney2005},
resulting either in an amplification or a decrease of the state weight.

For a degenerate system, $e^{i\varphi_{B}}$ is replaced by a unitary
matrix ${\cal U}_{B}$ which can always be diagonalized in an orthonormal
basis. Under the adiabatic evolution, each basis state then simply
gets multiplied by a phase factor. For an open system, the \emph{averaged
Berry matrix}, $\bar{{\cal U}}_{B}$, cannot be diagonalized in a
single orthonormal basis anymore since the eigenbasis of ${\cal U}_{B}$
depends on the respective trajectory realization. However, diagonalization
of $\bar{{\cal U}}_{B}$ is possible with \emph{two} orthonormal bases,
encoded by unitary matrices $u$ and $v$. Under the adiabatic evolution,
a vector in basis $v$ is mapped to a corresponding vector in $u$,
and multiplied by a real scaling factor $e^{-\Gamma_{g}}$ which depends
on the specific basis vector. We emphasize that such state mappings
have to be understood with caution since one can in principle only
average density matrices and not states \citep{WeissBook}. However,
many experimental observables for NAGD detection can be directly expressed
in terms of the action of $\bar{{\cal U}}_{B}$ on an arbitrary initial
state, see Ref.~\citep{prblong} and below. We now observe that for
the time-reversed protocol, a vector in $u$ is mapped to a vector
in $v$ and multiplied with the factor $e^{+\Gamma_{g}}$. We thus
see that NAGD comes with a nontrivial matrix structure due to the
existence of two bases. The details of the experimental protocol ---
and not only the orientation sense as in the Abelian case --- will
then determine whether NAGD implies an amplification or a suppression
of coherences.

\emph{Non-Abelian Adiabatic Dynamics.---}We consider a Hamiltonian
$H({\bm{\lambda}})$ that depends on $d$ classical parameters, $\lambda^{\mu}$,
forming the vector ${\bm{\lambda}}=(\lambda^{1},\ldots,\lambda^{d})$.
Suppose that for all relevant values of the parameters $\bm{\lambda}$,
the Hamiltonian spectrum consists of $M$ blocks with energy $E_{j}(\bm{\lambda})$
and degeneracy $N_{j}$. With a unitary, $U({\bm{\lambda}})$, and
a diagonal matrix, $D({\bm{\lambda}})=\sum_{j=1}^{M}E_{j}(\bm{\lambda})P_{j}$,
the Hamiltonian is given by $H({\bm{\lambda}})=U({\bm{\lambda}})D({\bm{\lambda}})U^{\dagger}({\bm{\lambda}})$,
where the $P_{j}$ are diagonal matrices projecting to the respective
block, $P_{j}P_{k}=\delta_{jk}P_{j}$. For now we focus on block $j=1$
with degeneracy $N\equiv N_{1}$. By imposing a parameter protocol
${\bm{\lambda}}(t)$, we obtain a time-dependent Hamiltonian $H(t)$.
We study closed loops, ${\bm{\lambda}}(0)={\bm{\lambda}}(T)$, and
without loss of generality assume $U({\bm{\lambda}}(0))=1$. For $N=1$,
one has a non-degenerate system with an Abelian Berry phase, while
for $N\geq2$ one obtains the non-Abelian Berry matrix \citep{Wilczek1984,Zee1988}.

Starting from an initial state in the $N$-fold degenerate subspace,
$|\psi(0)\rangle=|\psi_{0}\rangle$, the Schr\"odinger equation, $i\partial_{t}|\psi(t)\rangle=H(t)|\psi(t)\rangle$,
is solved by first transforming to the instantaneous eigenbasis, $|\psi(t)\rangle=U({\bm{\lambda}}(t))|\tilde{\psi}(t)\rangle$.
For sufficiently slow protocol ${\bm{\lambda}}(t)$, the adiabatic
theorem implies that the state must remain in the degenerate subspace
at all times. The final state follows as $|\tilde{\psi}(T)\rangle={\cal U}|\tilde{\psi}(0)\rangle$,
with ${\cal U}=e^{-i\int_{0}^{T}dtE_{1}}\,{\cal U}_{B}$ and the $N\times N$
Berry matrix \citep{Wilczek1984,Zee1988}
\begin{equation}
{\cal U}_{B}={\cal T}e^{-\int_{0}^{T}dt\ \dot{\lambda}^{\mu}(t)A_{\mu}({\bm{\lambda}}(t))}={\cal P}e^{-\oint d\lambda^{\mu}A_{\mu}}.\label{udef}
\end{equation}
Summation over repeated indices is always implied, ${\cal T}$ (${\cal P}$)
denotes time (path) ordering, and with the shorthand $\partial_{\mu}=\partial_{\lambda^{\mu}}$,
the non-Abelian Berry connection is given by \citep{Nakahara}
\begin{equation}
A_{\mu}({\bm{\lambda}})=P_{1}U^{\dagger}({\bm{\lambda}})\partial_{\mu}U({\bm{\lambda}})P_{1}=-A_{\mu}^{\dagger}({\bm{\lambda}}).\label{adef}
\end{equation}
As a path-ordered Wilson loop amplitude, ${\cal U}_{B}$ evidently
is of purely geometric origin.

\emph{Coupling to environment.---}We now allow for weak fluctuations
of the control parameters, ${\bm{\lambda}}(t)\to{\bm{\lambda}}(t)+\delta{\bm{\lambda}}(t)$,
around a base trajectory ${\bm{\lambda}}(t)$ \citep{foot1}. Here
the noise trajectory $\delta{\bm{\lambda}}(t)$ can be generated from
classical fluctuations of the control parameters and/or from a coupling
of the system to a quantum bath. In
the latter case, $\delta{\bm{\lambda}}(t)$ represents operators acting
on the bath Hilbert space which entangle the system with the bath and result in a non-unitary time evolution of the system. To simplify the analysis, we treat typical
system-bath couplings in Markov-Born approximation \citep{WeissBook},
where one obtains Gaussian statistics with vanishing mean, $\langle\delta\lambda^{\mu}(t)\rangle=0$,
and the correlation function
\begin{equation}
\left\langle \delta\lambda^{\mu}(t)\delta\lambda^{\nu}(t')\right\rangle =\sigma^{\mu\nu}\delta_{\tau_{c}}(t-t').\label{noisedef}
\end{equation}
The real positive $d\times d$ matrix with components $\sigma^{\mu\nu}=\sigma^{\nu\mu}$
contains the noise amplitudes and $\delta_{\tau_{c}}(t-t')$ is a
$\delta$-function broadened on the scale of the noise correlation
time $\tau_{c}$ \citep{foot2}. We note in passing that artificially
generated classical noise with such properties has been employed experimentally
for the Abelian case \citep{Berger2015}, allowing for in-detail investigations
of noise-related effects. In what follows, we denote the typical size
of $\sigma^{\mu\nu}$, e.g., the largest eigenvalue, by $\sigma$.
For simplicity, we assume that $\lambda^{\mu}$ (and thus also $\sigma^{\mu\nu}$
and $\sigma$) carries energy units. Below we impose three conditions:
(i) The noise correlation time $\tau_{c}$ is short against the protocol
duration $T$. (ii) The evolution does not mix block 1 with other
blocks. With the minimal energy difference ${\cal E}$ between $E_{1}$
and other eigenenergies, this implies ${\cal E}\tau_{c}\gg1$. (iii)
The system-bath coupling is weak, $\sigma\ll{\cal E}$. In summary,
we arrive at the inequality chain
\begin{equation}
{\cal E}T\gg{\cal E}\tau_{c}\gg1\gg\sigma/{\cal E}.\label{conditions}
\end{equation}

\emph{Averaged Berry matrix.---}Consider first a single realization
of the control parameter trajectory, ${\bm{\lambda}}(t)+\delta{\bm{\lambda}}(t)$.
Expanding Eq.~\eqref{udef} in powers of $\delta\lambda^{\mu}$ and
using Eq.~\eqref{conditions}, we obtain ${\cal U}=e^{-i\int_{0}^{T}dtE_{1}\left({\bm{\lambda}}(t)+\delta{\bm{\lambda}}(t)\right)}\,{\cal U}_{{B}},$
with the Berry matrix
\begin{equation}
{\cal U}_{B}=\mathcal{P}\exp{\oint d\lambda^{\mu}\left[-A_{\mu}+\delta\lambda^{\nu}F_{\mu\nu}+{\cal O}(\delta\lambda^{2})\right]},\label{Udeltalambda}
\end{equation}
and the non-Abelian Berry curvature (or field strength) tensor \citep{Nakahara},
\begin{equation}
F_{\mu\nu}^ {}({\bm{\lambda}})=\partial_{\mu}A_{\nu}-\partial_{\nu}A_{\mu}+[A_{\mu},A_{\nu}]=-F_{\mu\nu}^{\dagger}.\label{fmunu}
\end{equation}
We next perform the Gaussian average over the fluctuations $\delta\lambda^{\mu}(t)$
according to Eq.~\eqref{noisedef}. We then obtain
\begin{equation}
\bar{{\cal U}}=e^{-i\int_{0}^{T}dtE_{1}}e^{-\frac{1}{2}\int_{0}^{T}dt\ \sigma^{\mu\nu}\partial_{\mu}E_{1}\partial_{\nu}E_{1}}\,\bar{{\cal U}}_{B},\label{dynamic}
\end{equation}
where the averaged Berry matrix can again be expressed as a path-ordered
exponential,
\begin{equation}
\bar{{\cal U}}_{B}=\mathcal{P}\exp\oint d\lambda^{\mu}\left(-A_{\mu}+i\sigma^{\nu\rho}F_{\nu\mu}\partial_{\rho}E_{1}\right),\label{avU}
\end{equation}
and thus also represents a geometric contribution. The terms dropped
in the exponent of Eq.~\eqref{avU} are of order ${\cal O}\left(\frac{1}{\mathcal{E}T},\frac{\sigma^{2}}{\mathcal{E}^{2}},\frac{\sigma}{\mathcal{E}^{2}\tau_{c}}\right)$
\citep{prblong} and thus vanish according to Eq.~\eqref{conditions}.
The first term in Eq.~\eqref{dynamic} contains the dynamic phase.
The second term describes dynamic dephasing, with a trivial matrix
structure in the $N$-dimensional Hilbert space of block 1
and the exponent $\Gamma_{{\rm dyn}}T\sim\sigma T$. Note that this
term stays invariant under the time reversal of the protocol, ${\bm{\lambda}}'(t)={\bm{\lambda}}(T-t)$.
The nontrivial matrix structure of $\bar{{\cal U}}$ is encoded by
the averaged Berry matrix $\bar{{\cal U}}_{B}$ in Eq.~\eqref{avU},
which contains both the non-Abelian Berry phase of the base path and
an extra piece from the interplay of dynamic ($\sim\partial_{\rho}E_{1}$)
and geometric ($\sim F_{\nu\mu}$) phase fluctuations. Noting that
$(iF_{\nu\mu})^{\dagger}=iF_{\nu\mu}$, this term yields a Hermitian
contribution from Eq.~\eqref{avU} which is responsible for geometric
dephasing.
We stress that for classical fluctuations $\delta{\bm\lambda}$, a non-unitary matrix $\cal U_B$ emerges only after
performing an average over fluctuation realizations.
Finally, since replacing $\bm{\lambda}(t)\rightarrow\bm{\lambda}'(t)$
reverses the path ordering and flips the sign of $d\lambda^{\mu}$,
 the time-reversed protocol has the averaged Berry
matrix $\bar{{\cal U}}_{B}^{-1}$.

\emph{Polar decomposition.---}The non-unitary $N\times N$ matrix
$\bar{{\cal U}}_{B}$ admits the singular value decomposition $\bar{{\cal U}}_{B}=u\Lambda v^{\dagger}$
\citep{Horn}, where the unitaries $u$ and $v$ encode the two bases
introduced above. The diagonal matrix $\Lambda$ describes NAGD and
contains the respective real scaling factors $e^{-\Gamma_{g}}$. We
thus have the polar decomposition \citep{Horn}
\begin{equation}
\bar{{\cal U}}_{B}=VR,\quad V=uv^{\dagger},\quad R=v\Lambda v^{\dagger}=R^{\dagger},\label{polardecomp}
\end{equation}
where $V$ is a unitary rotation and $R$ is a positive semi-definite
Hermitian matrix. For generic non-Abelian systems, one has $[V,R]\ne0$.
We will see below that the non-commutativity of $V$ and $R$ has
profound and experimentally observable consequences. One can diagonalize
the unitary rotation, $V=w\Phi w^{\dagger}$, with a unitary $w$
and a diagonal matrix $\Phi$ containing phase factors. Hence $\bar{{\cal U}}_{B}$
is composed of (i) a real scaling transformation ($\Lambda$) applied
in the basis encoded by $v$, followed by (ii) a phase multiplication
($\Phi$) in the basis encoded by $w$. We next observe that the averaged
Berry matrix for the reversed protocol, $\bar{{\cal U}}_{B}^{-1}$,
has the polar decomposition
\begin{equation}
\bar{{\cal U}}_{B}^{-1}=\tilde{V}\tilde{R},\quad\tilde{V}=V^{\dagger},\quad\tilde{R}=u\Lambda^{-1}u^{\dagger}=\tilde{R}^{\dagger}.\label{polarback}
\end{equation}
While the unitary rotation is simply $\tilde{V}=V^{\dagger}$, the
Hermitian matrix $\tilde{R}$ contains the diagonal matrix $\Lambda^{-1}$
(characteristic of geometric dephasing) but in a different basis than
for the forward direction, cf.~Eq.~\eqref{polardecomp}. The matrix
structure of NAGD thus plays a crucial role when comparing results
for different directions of the protocol.

\begin{figure}
\centering \includegraphics[width=0.95\columnwidth]{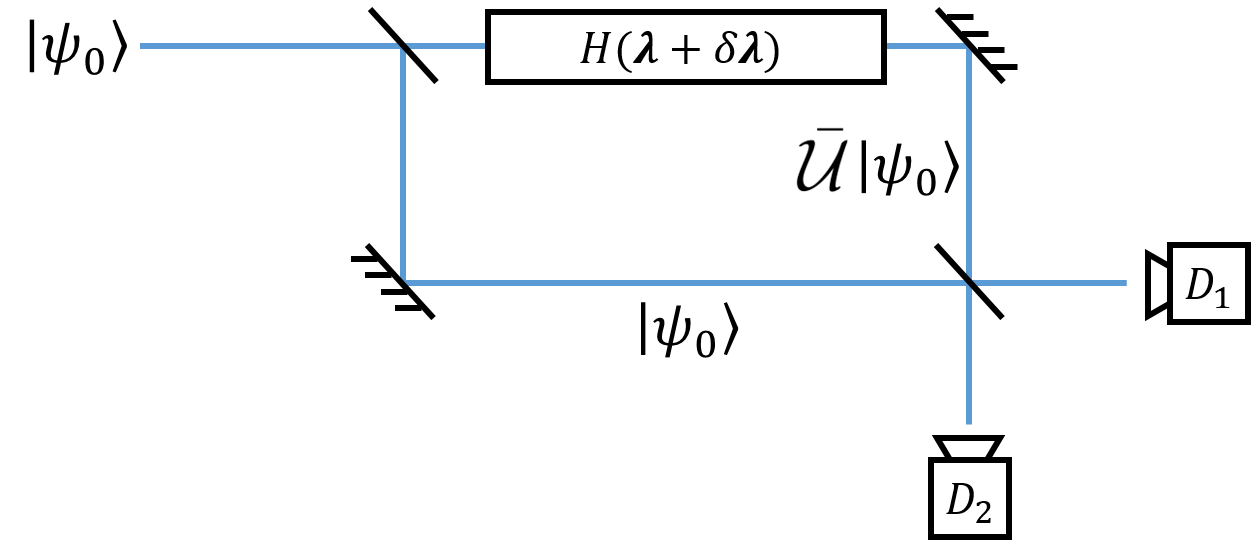} \caption{\label{fig1} An interferometric setup for observing NAGD. For details,
see main text.}
\end{figure}

\emph{NAGD detection by Mach-Zehnder interferometry.---}The conceptually
simplest setup to probe $\bar{{\cal U}}_{B}$ is the Mach-Zehnder
interferometer sketched in Fig.~\ref{fig1}. Consider a particle
with an internal spin-$S$ degree of freedom (where $S\ge1$), which
is prepared in the initial state $|\psi(0)\rangle=|\psi_{0}\rangle$.
In one arm of the interferometer, the spin dynamics evolves according
to the Hamiltonian $H({\bm{\lambda}}(t)+\delta{\bm{\lambda}}(t))$,
while in the other arm, $H=0$. The time of flight is then given by
$T$. Averaging over the parameter fluctuations $\delta{\bm{\lambda}}$,
i.e., over many experimental runs, the probability for the particle
to appear at the respective detector $D_{1,2}$ in Fig.~\ref{fig1}
is given by $\bar{P}_{1,2}=(1\pm\mathrm{Re}\langle\psi_{0}|\bar{\mathcal{U}}|\psi_{0}\rangle)/2$,
with $\bar{\mathcal{U}}$ in Eqs.~\eqref{dynamic} and \eqref{avU}.
It stands to reason that by repeating such an experiment for many
different initial states $|\psi_{0}\rangle$, the full matrix structure
of $\bar{\mathcal{U}}$, and thus of $\bar{{\cal U}}_{B}$, can be
determined. Natural candidates for such experiments are given by $S=3/2$
nuclei experiencing nuclear quadrupole resonance \citep{Tycko1987,Zwanziger1990},
where Berry connection and curvature expressions can be found in Ref.~\citep{Zee1988}.

\emph{Two-block interference for NAGD detection.---}In experiments
on condensed-matter systems featuring Abelian \citep{Leek2007,Berger2015}
or non-Abelian Berry phases \citep{Hasan2010,Yang2014,Murakami2003,AbdumalikovJr2013},
it is usually not possible to probe spatial superpositions as shown
in Fig.~\ref{fig1}. Fortunately, in such cases, one may employ a
different interference experiment as described next. Our interference
scheme takes into account two blocks ($j=1,2$), with the respective
energy $E_{j}({\bm{\lambda}})$, degeneracy $N_{j}$, and projector
$P_{j}$. Let us start from an arbitrary initial state within the
two blocks, $(P_{1}+P_{2})|\psi(0)\rangle=|\psi(0)\rangle$, and run
a parameter loop protocol with Hamiltonian $H({\bm{\lambda}}(t)+\delta{\bm{\lambda}}(t))$
as before. At time $t=T$, the expectation value of an operator of
the form ${\bm{M}}=P_{1}{\bm{M}}^{(12)}P_{2}+{\rm H.c.}$ is measured.
(Block-diagonal contributions, e.g., $P_{1}{\bm{M}}^{(11)}P_{1}$,
do not cause NAGD signatures \citep{prblong}.) For a given parameter
trajectory, we thus obtain ${\cal M}=\langle\psi(T)|{\bm{M}}|\psi(T)\rangle$.
Averaging this measurement over fluctuations, the result can be cast
as the expectation value of an operator $\bar{\bm{M}}$ in the \emph{known}
initial state,
\begin{equation}
\bar{{\cal M}}=\langle\psi(0)|\bar{\bm{M}}|\psi(0)\rangle,\quad\bar{\bm{M}}=\bigl(\bar{{\cal U}}^{(1)}\bigr)^{\dagger}{\bm{M}}^{(12)}\bar{{\cal U}}^{(2)}+{\rm H.c.},\label{interferometry}
\end{equation}
with $\bar{{\cal U}}^{(j)}=e^{-i\int_{0}^{T}dt\left[E_{j}-\frac{i}{2}\sigma^{\mu\nu}\partial_{\mu}E_{j}\partial_{\nu}\tilde{E}_{j}\right]}\,\bar{{\cal U}}_{B}^{(j)}.$
The averaged Berry matrices $\bar{{\cal U}}_{B}^{(j)}$ are defined
as in Eq.~\eqref{avU} but with the Berry connection $A_{\mu}^{(j)}$
and curvature $F_{\nu\mu}^{(j)}$ of the respective block, see Eqs.~\eqref{adef}
and \eqref{fmunu}, and using $E_{1}\to\tilde{E}_{j}$ with $\tilde{E}_{1}=-\tilde{E}_{2}=E_{1}-E_{2}$.
Importantly, the operator $\bar{\bm{M}}$ in Eq.~\eqref{interferometry}
contains the averaged Berry matrices. The above protocol thus offers
experimental access to the physics associated with NAGD. In particular,
by systematic variation of the initial state $|\psi(0)\rangle$ and
of the measured operator ${\bm{M}}$, one can map out the averaged
Berry matrices. We emphasize that such a detection scheme relies on
interference between two different blocks. Let us also remark that
for $N_{1}=N_{2}$, a powerful spin-echo type variant of this protocol
exists where dynamic phases drop out completely \citep{prblong}.
Below we discuss a concrete example for such a protocol using modified
Majorana braiding setups.

\begin{figure}
\centering \includegraphics[width=1\columnwidth]{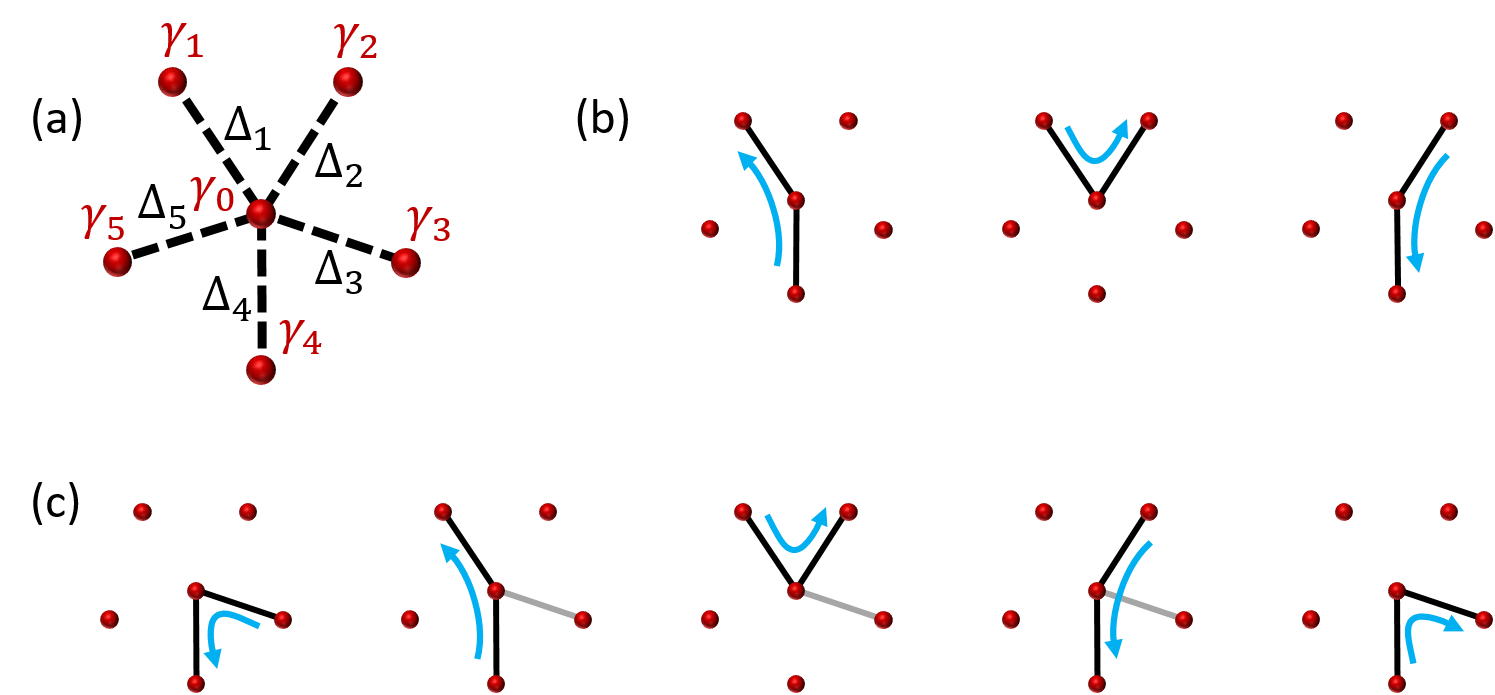} \caption{Schematic setup for NAGD detection using noisy Majorana braiding protocols.
(a) Five-star Majorana setup with tunnel couplings $\Delta_{j=1,\ldots,5}$
between Majorana operators $\gamma_{0}$ and $\gamma_{j}$. (b) Three-star
setup with $\Delta_{3}=\Delta_{5}=0$ (solid black lines indicate
$\Delta_{j}>0$) and topologically protected Majorana braiding protocol
\citep{Alicea2012,Sau2011,VanHeck2012}. The shown sequence implies
an exchange of $\gamma_{1}$ and $\gamma_{2}$. For instance, in the
first step, one starts with only $\Delta_{4}>0$. The blue arrow then
means that $\Delta_{4}(t)$ is slowly reduced to zero while $\Delta_{1}(t)$
is simultaneously ramped up. (c) Elementary steps for the NAGD detection
protocol, where $\Delta_{5}=0$ at all times. The grey solid line
indicates an additional coupling $\Delta_{3}'\protect\neq0$ needed
for generating NAGD. For a full description of the protocol, see main
text.}
\label{fig2}
\end{figure}

\emph{Noisy Majorana braiding setup.---}Let us next consider the
five-star Majorana setup in Fig.~\ref{fig2}(a) which we model by
the Hamiltonian $H_{M}(t)=i\gamma_{0}\sum_{j=1}^{5}\Delta_{j}(t)\gamma_{j}$,
where the real-valued tunnel couplings $\Delta_{j}$ correspond to
${\bm{\lambda}}(t)$. They can be tuned, e.g., by electric gates.
The Majorana operators $\gamma_{k}^ {}=\gamma_{k}^{\dagger}$ satisfy
the anticommutator algebra $\{\gamma_{k},\gamma_{l}\}=2\delta_{kl}$
\citep{Alicea2012}. Since total fermion number parity is conserved,
we choose, say, the even parity sector, corresponding to a four-dimensional
Hilbert space. $H_{M}$ has the two-fold degenerate energy levels
$E_{\pm}=\pm{\cal E}/2$ with ${\cal E}=2\sqrt{\sum_{j}\Delta_{j}^{2}}$,
corresponding to $E_{1}$ and $E_{2}$ above. We assume that the fluctuations
$\delta\Delta_{j}(t)$ are uncorrelated for different tunnel links.
Since tunnel couplings $\Delta_{j}$ are exponentially sensitive to
fluctuations, the latter act in a multiplicative way, i.e., $\delta\Delta_{j}(t)\sim\Delta_{j}(t)$
\citep{Karzig2016,Rahmani2017}, and Eq.~\eqref{noisedef} gives
\begin{equation}
\langle\delta\Delta_{j}(t)\delta\Delta_{k}(t')\rangle=\kappa_{j}\Delta_{j}^{2}(t)\delta_{jk}\delta_{\tau_{c}}(t-t'),\label{noiseM}
\end{equation}
where Eq.~\eqref{conditions} implies $\kappa_{j}{\cal E}\ll1$ for
the noise amplitudes $\kappa_{j}$. We assume below that only $\kappa_{4}\ne0$,
but see Ref.~\citep{prblong} for the general case.

Let us now recall that Majorana braiding is conventionally discussed
for a three-star setup \citep{Alicea2012,Sau2011,VanHeck2012,Cheng2011,Karzig2016,Knapp2016,Rahmani2017}.
The corresponding braiding protocol is shown in Fig.~\ref{fig2}(b),
where at all times only two $\Delta_{j}(t)$ are non-zero. As a consequence,
even when including the noise in Eq.~\eqref{noiseM}, the \emph{geometric}
trajectory in the relevant parameter space $\{\Delta_{j}(t)/{\cal E}\}$
does not fluctuate. According to Eq.~\eqref{avU}, NAGD is due to
cross-correlations of energy fluctuations and geometric trajectory
fluctuations. The absence of the latter thus implies the absence of
NAGD, reflecting the topological protection of this braiding scheme
\citep{prblong}. By contrast, for the same protocol, the presence
of an extra coupling ($\Delta_{3}'\ne0$) removes the protection and
allows for geometric trajectory fluctuations. We thereby obtain an
appealing candidate for NAGD detection. The sequence in Fig.~\ref{fig2}(c)
provides an example, where during intermediate steps, we set $\Delta_{3}(t)=\Delta_{3}'\neq0$.
Even though $\Delta_{3}'$ does not fluctuate, geometric trajectory
fluctuations can then develop from $\delta\Delta_{4}(t)$ contributions.

\emph{Spin-echo protocol for the Majorana setup.---}We now discuss
the NAGD detection protocol depicted in Fig.~\ref{fig2}(c). We start
from the initial state $|\psi_{0}\rangle\sim|0_{12}0_{03}0_{45}\rangle+|1_{12}1_{03}0_{45}\rangle$,
where $|n_{jk}\rangle$ with $n_{jk}=(p_{jk}+1)/2=0,1$ encodes the
eigenvalue $p_{jk}=\pm1$ of the Majorana parity operator $\hat{p}_{jk}=i\gamma_{j}\gamma_{k}$.
We note that $|\psi_{0}\rangle$ can be prepared from $|0_{12}0_{03}0_{45}\rangle$
by measuring the parity operator $\hat{p}_{02}$ with outcome $p_{02}=-1$
\citep{Plugge2017}. Readout of Majorana parities can be performed
as described in Refs.~\citep{prblong,Fu2010,Flensberg2011,Aasen2016,Plugge2016,Plugge2017,Karzig2017}.
One then runs (i) the sequence shown in Fig.~\ref{fig2}(c), followed
by (ii) the time-reversed sequence but with $\Delta'_{3}=0$. At this
point ($t=T$), (iii) one applies the flip operator $\hat{p}_{02}=i\gamma_{0}\gamma_{2}$
which will effectively exchange both blocks. Also for this operation,
implementation proposals are available \citep{Flensberg2011,Aasen2016,Plugge2016,Plugge2017,Karzig2017}.
We then, (iv) and (v), simply repeat steps (i) and (ii). The protocol
ends by (vi) applying the flip operator $\hat{p}_{02}$ again. The
Majorana parities $\hat{p}_{0j}=i\gamma_{0}\gamma_{j}$ are then measured
in the final state, and the results are subsequently averaged over
the noise to yield the expectation values $\bar{p}_{0j}$. Fortunately,
this problem is simple enough to admit analytical predictions \citep{prblong}.
We can thus avoid a full-fledged numerical analysis of path-ordered
expressions for the averaged Berry matrices which is required in most
other applications. For overall orientation sense $\eta=\pm$ of the
protocol, using $\cos{\alpha}=2\Delta_{3}'/{\cal E}$ and $\beta=(\pi/4)\cos{\alpha}$,
we find
\begin{eqnarray}
\bar{p}_{01} & = & e^{-4\Gamma_{{\rm dyn}}T}{\cal O}(\zeta^{2}),\quad\bar{p}_{03}=\bar{p}_{05}=0,\nonumber \\
\bar{p}_{02} & = & e^{-4\Gamma_{{\rm dyn}}T}\left[-1+2\eta\zeta\sin(4\beta)+{\cal O}(\zeta^{2})\right],\\
\bar{p}_{04} & = & -8e^{-4\Gamma_{{\rm dyn}}T}\zeta\sin^{2}\beta+{\cal O}(\zeta^{2}),\nonumber
\end{eqnarray}
with a dynamic dephasing rate $\Gamma_{{\rm dyn}}\sim\kappa_{4}{\cal E}^{2}$
and the dimensionless NAGD parameter $\zeta=\frac{3\pi}{16}\kappa_{4}{\cal E}\sin^{4}\alpha\cos\alpha$.
While the contribution $\sim\zeta$ to $\bar{p}_{02}$ depends on
$\eta=\pm$ in the same manner as expected for Abelian geometric dephasing,
the average $\bar{p}_{04}$ vanishes without noise and otherwise does
\emph{not} change sign under $\eta\to-\eta$. This last feature, in
particular, represents compelling evidence for NAGD \citep{prblong}.

\emph{Conclusions.---}We have shown that the adiabatic quantum dynamics
within a degenerate subspace of an open system will contain universal
NAGD contributions. The unique matrix structure of NAGD allows for
its clear-cut identification using interference experiments. Interesting
directions for future work include extensions to other topological
entities (e.g., parafermions or non-Abelian quasi-particles in quantum
Hall setups), applications to weak anti-localization in condensed-matter
systems with a non-Abelian Berry connection, as well as the study
of non-adiabatic corrections
which can be identified by the fact that other blocks become populated.

\begin{acknowledgements} We thank A. Altland for discussions. This
project has been funded by the Deutsche Forschungsgemeinschaft (DFG,
German Research Foundation), Projektnummer 277101999, TRR 183 (project
C01). \end{acknowledgements}


\begin{thebibliography}{10}
\bibitem{foot1} We assume $\delta{\bm{\lambda}}(t=0,T)=0$. As long
as Eq.~\eqref{conditions} holds, however, this is not essential.
\end{thebibliography}
\end{document}